\documentclass{elsart}

\begin{document}

\begin{frontmatter}

\title{An entanglement measure based on the capacity of dense coding}
\author{Tohya Hiroshima}
\ead{tohya@frl.cl.nec.co.jp}
\address{Fundamental Research Laboratories, NEC Corporation,\\
34 Miyukigaoka, Tsukuba, 305-8501, Japan,}

\begin{abstract}
An asymptotic entanglement measure for any bipartite states is derived in
the light of the dense coding capacity optimized with respect to local
quantum operations and classical communications. General properties and some
examples with explicit forms of this entanglement measure are investigated.
\end{abstract}


\end{frontmatter}

Quantum entanglement is one of the key ingredients in various types of
quantum information processing. A notable example is dense coding \cite{BW},
which doubles the capacity of transmission of classical information assisted
by an maximally entangled pair of qubits shared between the sender (Alice)
and receiver (Bob). Several authors have studied the capacity of dense
coding in various situations \cite{BSST,BPV,Hiroshima,Bowen,HHHLT,Winter}. In this
paper, the author derives an entanglement measure for any bipartite states
in the light of the capacity of dense coding and investigates its properties
systematically. First, the general scheme of dense coding with a mixed state
on the Hilbert space $C^{d}\otimes C^{d}$ is described. Alice performs one
of the local unitary transformations $U_{i}\in U(d)$ on her $d$-dimensional
quantum system in order to put the initially shared entangled state $\rho $
in $\rho _{i}=(U_{i}\otimes {\bf I}_{d})\rho (U_{i}^{\dagger }\otimes {\bf I}%
_{d})$ with {\it a priori} probability $p_{i}$ $(i=1,2,\cdots ,i_{\max })$,
and then she sends her quantum system to Bob. Upon receiving this quantum
system, Bob performs a suitable measurement on $\rho _{i}$ to extract the
signal. The optimal amount of information that can be conveyed is known to
be bounded from above by the Holevo quantity \cite{Kholevo}, 
\begin{equation}
\chi =S(\overline{\rho })-\sum_{i=1}^{i_{\max }}p_{i}S(\rho _{i}),
\label{eq:Holevo}
\end{equation}
where $S(\rho )=-{\rm Tr}(\rho \log _{2}\rho )$ denotes the von Neumann
entropy and $\overline{\rho }=\sum_{i=1}^{i_{\max }}p_{i}\rho _{i}$ is the
average density matrix of the signal ensemble. Since the Holevo quantity is
asymptotically achievable \cite{Holevo,SW}, Eq.~(\ref{eq:Holevo}) is used
here as the definition of the capacity of dense coding. Capacity $\chi $ is
maximized for signal states $\rho _{i}$ with mutually orthogonal unitary
transformations, ${\rm Tr}\left( U_{i}^{\dagger }U_{j}\right) =d\delta _{ij}$
and equal probabilities $p_{i}=d^{-2}$ ($i_{\max }=d^{2}$) \cite{Hiroshima}.
The optimal capacity is written as $\chi ^{*}(\rho )=\log _{2}d+I_{B}(\rho )$%
, where $I_{B}(\rho )=S(\rho ^{B})-S(\rho )$ is the coherent information
with $\rho ^{B}={\rm Tr}_{A}\rho $. Since $\max \left[ S(\rho ^{A})-S(\rho
),S(\rho ^{B})-S(\rho )\right] \leq E_{R}(\rho )$ \cite{PVP}, 
\begin{equation}
I_{B}(\rho )\leq E_{R}(\rho ),  \label{eq:IBER}
\end{equation}
and the capacity $\chi ^{*}(\rho )$ satisfies $\chi ^{*}(\rho )\leq \log
_{2}d+E_{R}(\rho )$ \cite{Hiroshima}. Here, $E_{R}(\rho )$ is the relative
entropy of entanglement \cite{VPRK,VP} for states $\rho $ defined as $%
E_{R}(\rho )=\min_{\sigma \in {\mathcal{D}}}S(\rho ||\sigma )$, where ${\mathcal{D}}$
the set of states with positive partial transpose (PPT states) and $S(\rho
||\sigma )={\rm Tr}\left[ \rho \left( \log _{2}\rho -\log _{2}\sigma \right)
\right] $ is the quantum relative entropy of $\rho $ with respect to $\sigma 
$.

Note that this capacity is optimal in the sense that Alice and Bob uses a
given mixed state $\rho $ as a resource for dense coding without any
changes. If they are allowed to perform local quantum operations and
classical communications (LQCC) on the initially shared mixed state $\rho $
prior to the dense coding, the capacity could be enhanced further. The
maximally achievable capacity thus obtained \cite{COMMENT} is 
\begin{equation}
\chi _{\max }^{*}(\rho ) =\log _{2}d+\lim_{n\rightarrow \infty
}\sup_{\Lambda _{n}}\frac{1}{n}I_{B}\left( \Lambda _{n}(\rho ^{\otimes
n})\right)
\equiv \log _{2}d+E_{dc}(\rho ). \label{eq:Edc}
\end{equation}
Namely, $E_{dc}(\rho )$ is the asymptotic limit of the achievable
(normalized) coherent information over the sequence of LQCC operations $%
\left\{ \Lambda _{n}\right\} $ or an LQCC protocol.

Hereafter the properties of $E_{dc}(\rho )$ defined in Eq.~(\ref{eq:Edc})
are examined. $E_{dc}(\rho )$ is the maximal dense coding capacity
subtracted by the classically achievable capacity $\log _{2}d$; it
represents the maximal contribution of entanglement to the dense coding
capacity. As shown in the following, $E_{dc}(\rho )$ is a measure of
entanglement of $\rho $. Before proving this, the following inequalities
must be proved. 
\begin{equation}
E_{D}(\rho )\leq E_{dc}(\rho )\leq E_{R}^{\infty }(\rho ),
\label{eq:EDEdcERinf}
\end{equation}
where $E_{D}(\rho )$ and $E_{R}^{\infty }(\rho )$ are, respectively, the
distillable entanglement \cite{BDSW} and the asymptotic relative entropy of
entanglement \cite{AEJPVD} of $\rho $, both of which are asymptotic
entanglement measures. $E_{R}^{\infty }(\rho )$ is defined as the average
relative entropy of entanglement per copy: 
\begin{equation}
E_{R}^{\infty }(\rho )=\lim_{n\rightarrow \infty }\frac{E_{R}(\rho ^{\otimes
n})}{n}.
\end{equation}
Noting the subadditivity of the relative entropy of entanglement, i.e., $%
E_{R}(\rho ^{\otimes n})\leq nE_{R}(\rho )$ \cite{VPRK}, a weaker version of
Eq.~(\ref{eq:EDEdcERinf}) is obtained: 
\begin{equation}
E_{D}(\rho )\leq E_{dc}(\rho )\leq E_{R}(\rho ).  \label{eq:EDEdcER}
\end{equation}
Although the proof of the first part of Eq.~(\ref{eq:EDEdcERinf}) is
essentially the same as that in \cite{HHHa}, the proof is described here for
completeness. It is always possible to consider that the distillation
protocol is ended by $U\otimes U^{*}$ twirling \cite{HH} so that the final
state is an isotropic state of the form 
\begin{equation}
\rho \left( F_{n},d_{n}\right) =pP_{+}(C^{d_{n}})+(1-p)\frac{1}{d_{n}^{2}}%
{\bf I}_{d_{n}},
\end{equation}
where $F_{n}={\rm Tr}\left[ \rho (F_{n},d_{n})P_{+}(C^{d_{n}})\right] $ is
the fidelity, ${\bf I}_{d_{n}}$ is the identity of dimensions $d_{n}$, and 
\begin{equation}
P_{+}(C^{d})=\left| \psi _{+}(C^{d})\right\rangle \left\langle \psi
_{+}(C^{d})\right|   \label{eq:Projector}
\end{equation}
is the maximally entangled state of a $C^{d}\otimes C^{d}$ system. In Eq.~(%
\ref{eq:Projector}), $\left| \psi _{+}(C^{d})\right\rangle =\frac{1}{\sqrt{d}%
}\sum_{i=1}^{d}\left| ii\right\rangle $ with $\left| i\right\rangle $ are
basis vectors in $C^{d}$. Because the protocol mentioned above is not
necessarily optimal for $E_{dc}$, $E_{dc}(\rho )\geq \lim_{n\rightarrow
\infty }\frac{1}{n}I_{B}\left( \rho \left( F_{n},d_{n}\right) \right) $. The
coherent information for $\rho \left( F_{n},d_{n}\right) $ is easily
calculated as 
\begin{equation}
I_{B}\left( \rho \left( F_{n},d_{n}\right) \right)  =\log
_{2}d_{n}+F_{n}\log _{2}F_{n}
+(1-F_{n})\log _{2}\frac{1-F_{n}}{d_{n}^{2}-1}.
\end{equation}
By definition of the distillable entanglement \cite{BDSW}, $F_{n}\rightarrow
1$ and $\frac{\log _{2}d_{n}}{n}\rightarrow E_{D}(\rho )$ for $n\rightarrow
\infty $. Therefore, $E_{dc}(\rho )\geq E_{D}(\rho )$. The proof of the
second part of Eq.~(\ref{eq:EDEdcERinf}) is as follows. Equation~(\ref
{eq:IBER}) and the weak monotonicity (see below) of $E_{R}$ \cite{VP} give 
\begin{equation}
E_{dc}(\rho )\leq \lim_{n\rightarrow \infty }\sup_{\Lambda _{n}}\frac{1}{n}%
E_{R}\left( \Lambda _{n}(\rho ^{\otimes n})\right) \leq \lim_{n\rightarrow
\infty }\frac{1}{n}E_{R}(\rho ^{\otimes n}).
\end{equation}
The right-hand side is, by definition, $E_{R}^{\infty }(\rho )$. Therefore, $%
E_{dc}(\rho )\leq E_{R}^{\infty }(\rho )$.

The quantity thus defined is an entanglement measure; namely, it satisfies
the following properties \cite{HHHa,Horodecki,DHR}.

\begin{itemize}
\item[(i)]  $E_{dc}(\rho )=0$ for any separable state $\rho $.

\item[(ii)]  $E_{dc}(\rho )\geq 0$.

\item[(iii)]  For a pure state $\left| \phi \right\rangle \left\langle \phi
\right| $, $E_{dc}$ is the von Neumann entropy of the reduced density
matrix, e.g., the entropy of entanglement; 
\[
E_{dc}\left( \left| \phi \right\rangle \left\langle \phi \right| \right)
=S\left( {\rm Tr}_{A}\left( \left| \phi \right\rangle \left\langle \phi
\right| \right) \right) =S\left( {\rm Tr}_{B}\left( \left| \phi
\right\rangle \left\langle \phi \right| \right) \right) .
\]
In particular, $E_{dc}\left( P_{+}(C^{d})\right) =\log _{2}d$, where $%
P_{+}(C^{d})$ is the maximally entangled state of a $C^{d}\otimes C^{d}$
system [Eq.~(\ref{eq:Projector})].

\item[(iv)]  Partial additivity: $E_{dc}(\rho ^{\otimes n})=nE_{dc}(\rho )$.

\item[(v)]  Weak monotonicity: $E_{dc}\left( \Lambda (\rho )\right) \leq $ $%
E_{dc}(\rho )$, where $\Lambda $ is an LQCC operation. This is the most
important property required of the entanglement measure.

\item[(vi)]  Convexity on pure state decomposition: 
\[
E_{dc}\left( \sum_{i}p_{i}\left| \phi _{i}\right\rangle \left\langle \phi
_{i}\right| \right) \leq \sum_{i}p_{i}E_{dc}\left( \left| \phi
_{i}\right\rangle \left\langle \phi _{i}\right| \right) ,
\]
with $\sum_{i}p_{i}=1$ and $p_{i}\geq 0$

\item[(vii)]  Weak continuity: For any sequence of the pure state $\left|
\psi _{n}\right\rangle $ and the mixed state $\rho _{n}$ of a system $%
C^{d_{n}}\otimes C^{d_{n}}$ such that $\left\| \rho _{n}-\left| \psi
_{n}\right\rangle \left\langle \psi _{n}\right| \right\| _{1}\rightarrow 0$
and $d_{n}\rightarrow \infty $ for $n\rightarrow \infty $, 
\[
\lim_{n\rightarrow \infty }\frac{E_{dc}(\rho _{n})-E_{dc}(\left| \psi
_{n}\right\rangle \left\langle \psi _{n}\right| )}{\log d_{n}}=0.
\]
Here, $\left\| A\right\| _{1}$ denotes the trace norm of $A$; $\left\|
A\right\| _{1}={\rm Tr}\sqrt{A^{\dagger }A}$.
\end{itemize}

Properties (i)-(iii) are obvious from Eq.~(\ref{eq:EDEdcER}). The proof of
property (iv) is as follows. 
\begin{eqnarray}
E_{dc}(\rho ^{\otimes m}) &=&\lim_{n\rightarrow \infty }\sup_{\Lambda _{nm}}%
\frac{1}{n}I_{B}\left( \Lambda _{nm}(\rho ^{\otimes nm})\right) \nonumber\\
&=&m\lim_{n^{\prime }\rightarrow \infty }\sup_{\Lambda _{n^{\prime }}}\frac{1%
}{n}I_{B}\left( \Lambda _{n^{\prime }}(\rho ^{\otimes n^{\prime }})\right) 
=mE_{dc}(\rho ).
\end{eqnarray}
Property (v) is obvious because $E_{dc}(\rho )$ is the optimized quantity
with respect to LQCC protocols by definition and the tensor product of an
LQCC operation is also an LQCC operation. Property (vi) follows from Eq.~(%
\ref{eq:EDEdcER}) and the fact that both $E_{dc}(\rho )$ and $E_{R}(\rho )$
coincide on pure states; 
\begin{eqnarray}
E_{dc}\left( \sum_{i}p_{i}\left| \phi _{i}\right\rangle \left\langle \phi
_{i}\right| \right)  &\leq &E_{R}\left( \sum_{i}p_{i}\left| \phi
_{i}\right\rangle \left\langle \phi _{i}\right| \right)   \nonumber \\
&\leq &\sum_{i}p_{i}E_{R}\left( \left| \phi _{i}\right\rangle \left\langle
\phi _{i}\right| \right)
=\sum_{i}p_{i}E_{dc}\left( \left| \phi _{i}\right\rangle \left\langle \phi
_{i}\right| \right) .
\end{eqnarray}
The proof of property (vii) is given as follows. Noting the fact that $I_{B}$%
, $E_{dc}$, and $E_{R}$ coincide on pure states, the inequalities, $%
I_{B}(\rho _{n})\leq E_{dc}(\rho _{n})\leq E_{R}(\rho _{n})$ give 
\begin{eqnarray}
I_{B}(\rho _{n})-I_{B}\left( \left| \psi _{n}\right\rangle \left\langle \psi
_{n}\right| \right)  &\leq &E_{dc}(\rho _{n})-E_{dc}\left( \left| \psi
_{n}\right\rangle \left\langle \psi _{n}\right| \right)   \nonumber \\
&\leq &E_{R}(\rho _{n})-E_{R}\left( \left| \psi _{n}\right\rangle
\left\langle \psi _{n}\right| \right) .  \label{eq:Continuity1}
\end{eqnarray}
Firstly, 
\begin{equation}
\lim_{n\rightarrow \infty }\frac{E_{R}(\rho _{n})-E_{R}(\left| \psi
_{n}\right\rangle \left\langle \psi _{n}\right| )}{\log d_{n}}=0,
\label{eq:Continuity2}
\end{equation}
because $E_{R}$ is continuous. Secondly, Fannes' inequality \cite{Fannes}, 
\begin{equation}
\left| S(\rho )-S(\sigma )\right|  \leq \left\| \rho -\sigma \right\|
_{1}\log _{2}\dim {\mathcal{H}}
+\eta \left( \left\| \rho -\sigma \right\| _{1}\right) ,\label{eq:Fannes}
\end{equation}
plays a key role. It holds for any two states $\rho $ and $\sigma $ acting
on the Hilbert space ${\mathcal{H}}$ provided that $\left\| \rho -\sigma \right\|
_{1}\leq 1/e$. In Eq.~(\ref{eq:Fannes}), $\eta (s)=-s\log _{2}s$. Noting the
fact that the partial trace does not increase the trace norm and $\eta (s)$
is a monotonically increasing function for $0\leq s\leq 1/e$, Fannes'
inequality [Eq.~(\ref{eq:Fannes})] gives 
\begin{eqnarray}
\left| I_{B}(\rho _{n})-I_{B}\left( \left| \psi _{n}\right\rangle
\left\langle \psi _{n}\right| \right) \right|  &\leq &3\left\| \rho
_{n}-\left| \psi _{n}\right\rangle \left\langle \psi _{n}\right| \right\|
_{1}\log _{2}d_{n}  \nonumber \\
&&+2\eta \left( \left\| \rho _{n}-\left| \psi _{n}\right\rangle \left\langle
\psi _{n}\right| \right\| _{1}\right) ,
\end{eqnarray}
Therefore, 
\begin{equation}
\lim_{n\rightarrow \infty }\frac{\left| I_{B}(\rho _{n})-I_{B}\left( \left|
\psi _{n}\right\rangle \left\langle \psi _{n}\right| \right) \right| }{\log
d_{n}}=0.  \label{eq:Continuity3}
\end{equation}
From Eqs.~(\ref{eq:Continuity1}), (\ref{eq:Continuity2}), and (\ref
{eq:Continuity3}), the following equation is obtained: 
\begin{equation}
\lim_{n\rightarrow \infty }\frac{E_{dc}(\rho _{n})-E_{dc}(\left| \psi
_{n}\right\rangle \left\langle \psi _{n}\right| )}{\log d_{n}}=0.
\end{equation}
Namely, $E_{dc}$ is weakly continuous.

In addition to properties (i)-(vii), $E_{dc}(\rho )$ exhibits
superadditivity. Namely, $E_{dc}(\rho \otimes \sigma )\geq E_{dc}(\rho
)+E_{dc}(\sigma )$. The proof is as follows. Because of the additivity of
the coherent information, 
\begin{equation}
E_{dc}(\rho )+E_{dc}(\sigma )
=\lim_{n\rightarrow \infty }\sup_{\Lambda _{n}^{\rho }\otimes \Lambda
_{n}^{\sigma }}\frac{1}{n}I_{B}\left( \left( \Lambda _{n}^{\rho }\otimes
\Lambda _{n}^{\sigma }\right) (\rho ^{\otimes n}\otimes \sigma ^{\otimes
n})\right) .
\end{equation}
Here, even if the protocol $\left\{ \Lambda _{n}^{\rho }\otimes \Lambda
_{n}^{\sigma }\right\} $ is optimized, it is not necessarily the optimal one
for $\left\{ (\rho \otimes \sigma )^{\otimes n}\right\} $. Therefore, 
\begin{equation}
E_{dc}(\rho )+E_{dc}(\sigma ) \leq \lim_{n\rightarrow \infty
}\sup_{\Lambda _{n}^{\rho \otimes \sigma }}\frac{1}{n}I_{B}\left( \Lambda
_{n}^{\rho \otimes \sigma }(\rho \otimes \sigma )^{\otimes n}\right)
=E_{dc}(\rho \otimes \sigma ).
\end{equation}
It is not clear at present if the equality (full additivity) holds. The
convexity of the general form, 
\begin{equation}
E_{dc}\left( \sum_{i}p_{i}\rho _{i}\right) \leq \sum_{i}p_{i}E_{dc}(\rho
_{i}),
\end{equation}
is also doubtful. However, it should be noted that the breakdown of the full
additivity and the general convexity is not a drawback; it is argued that
these two requirements are too strong for asymptotic entanglement measures 
\cite{HHHa,Horodecki}.

Although it is in general quite difficult to calculate $E_{dc}(\rho )$,
there are some special mixed states in which $E_{dc}(\rho )$ is obtained
explicitly.

{\it Example 1---}This is the example by Rains \cite{Rains00,Rains01}. It is
called the maximally correlated state of a $C^{d}\otimes C^{d}$ system, and
takes the form 
\begin{equation}
\rho =\sum_{i,j=1}^{d}\alpha _{ij}\left| ii\right\rangle \left\langle
jj\right| .
\end{equation}
The relative entropy of entanglement is calculated as 
\begin{equation}
E_{R}(\rho )=I_{B}(\rho )=H(\alpha _{11},\alpha _{22},\cdots )-S(\rho ),
\end{equation}
where $H(\alpha _{11},\alpha _{22},\cdots )$ denotes the Shannon entropy of
probability distribution $\left\{ \alpha _{ii}\right\} $. From $E_{R}(\rho
)=I_{B}(\rho )$, $E_{dc}(\rho )=E_{R}^{\infty }(\rho )=E_{R}(\rho )$, which
is proved as follows: 
\begin{eqnarray}
E_{R}(\rho )\leq E_{dc}(\rho ) &\leq &\lim_{n\rightarrow \infty
}\sup_{\Lambda _{n}}\frac{1}{n}E_{R}\left( \Lambda _{n}(\rho ^{\otimes
n})\right)   \nonumber \\
&\leq &\lim_{n\rightarrow \infty }\frac{1}{n}E_{R}(\rho ^{\otimes n})
=E_{R}^{\infty }(\rho )\leq E_{R}(\rho ).
\end{eqnarray}
The first inequality is obvious because $I_{B}(\rho )\leq E_{dc}(\rho )$.
The second inequality is a consequence of Eq.~(\ref{eq:IBER}), and the third
inequality follows from the weak monotonicity of $E_{R}$. The last
inequality is a result of the subadditivity of $E_{R}$. The optimal LQCC
operation for $E_{dc}$ is simply $\Lambda _{n}=I_{d^{n}}$. It has been shown
that $E_{R}(\rho )$ is exactly the PPT distillable entanglement (distillable
entanglement with respect to positive partial transpose operations \cite
{Rains99}). Since the set of LQCC operations is a subset of the set of PPT
operations, $E_{D}(\rho )$ is the lower bound on the PPT distillable
entanglement. Therefore, $E_{dc}(\rho )=E_{R}^{\infty }(\rho )=E_{R}(\rho
)=I_{B}(\rho )\geq E_{D}(\rho )$. When $d=2$, the maximally correlated state
is equivalent to a mixture of two Bell states (a Bell diagonal state of rank
two) if ${\rm Tr}_{A}\rho ={\bf I}_{2}/2$. For this state the hashing
protocol of distillation yields the value of $E_{R}(\rho )=1-S(\rho )$ for
the distillable entanglement \cite{BDSW} so that $E_{dc}(\rho )=E_{D}(\rho )$%
.

{\it Example 2}---This is the example by Eisert {\it et al.} \cite{EFPPW}.
Suppose that Alice and Bob share initially $N=2J$ ($J=1,2,\cdots $) pair of
qubits each in the same state $\left| \phi \right\rangle =\alpha \left|
00\right\rangle +\beta \left| 11\right\rangle $. Hereafter in this example $%
\alpha =\beta =1/\sqrt{2}$ is assumed for simplicity. After the complete
loss of the order of Bob's particles, the initially shared pure state $%
\left| \phi \right\rangle ^{\otimes N}$ becomes a mixed state of the form 
\begin{equation}
\rho =\sum_{j=0}^{J}\sum_{\alpha _{j},\beta _{j}=1}^{d_{j}}p_{j}\left| \psi
_{j}(\alpha _{j},\beta _{j})\right\rangle \left\langle \psi _{j}(\alpha
_{j},\beta _{j})\right| ,
\end{equation}
where 
\begin{equation}
\left| \psi _{j}(\alpha _{j},\beta _{j})\right\rangle =\frac{1}{\sqrt{2j+1}}%
\sum_{m=-j}^{j}\left| j,m,\alpha _{j}\right\rangle \left| j,m,\beta
_{j}\right\rangle ,
\end{equation}
$p_{j}=(2j+1)/(d_{j}2^{2J})$, and 
%$d_{j}=\frac{2j+1}{2J+1}\displaystyle 2J+1 \atopwithdelims()\displaystyle J-j$
$d_{j}=\frac{2j+1}{2J+1}{%
%TCIMACRO{\binom{2J+1}{J-j}}
%BeginExpansion
{2J+1 \choose J-j}%
%EndExpansion
}$ is the multiplicity of the $j$-representation in $SU(2)^{\otimes N}$. It
is easy to calculate the coherent information; 
\begin{equation}
I_{B}(\rho )=\sum_{j=0}^{J}d_{j}^{2}p_{j}\left[ \log _{2}(2j+1)-\log
_{2}d_{j}\right] .
\end{equation}
On the other hand, the relative entropy of entanglement and the distillable
entanglement are calculated as \cite{EFPPW} 
\begin{equation}
E_{R}(\rho )=E_{D}(\rho )=\sum_{j=0}^{J}d_{j}^{2}p_{j}\log _{2}(2j+1)
\end{equation}
so that $E_{dc}(\rho )=E_{R}^{\infty }(\rho )=E_{R}(\rho )=E_{D}(\rho )\geq
I_{B}(\rho )$. The first three equalities follow from Eqs.~(\ref
{eq:EDEdcERinf}) and (\ref{eq:EDEdcER}) and the subadditivity of $E_{R}$.
The last equality holds only for $J=1$ ($d_{0}=d_{1}=1$). The optimal
distillation is the optimal LQCC protocol for $E_{dc}$.

Two examples described above show that it is reasonable to conjecture that
the optimal protocol for $E_{dc}(\rho )$ is either the identity [$%
E_{dc}(\rho )=I_{B}(\rho )\geq E_{D}(\rho )]$ or the optimal distillation
protocols [$E_{dc}(\rho )=E_{D}(\rho )\geq I_{B}(\rho )]$. However,
Horodecki {\it et al.} have conjectured that $I_{B}(\rho )$ does not exceed
the one-way distillable entanglement (distillable entanglement with local
operations plus one-way classical communications) for any state $\rho $ \cite
{HHHb}. If this conjecture (hashing inequality) is true, 
\begin{equation}
E_{D}(\rho )\geq I_{B}(\rho )  \label{eq:hashing}
\end{equation}
holds for any state $\rho $. This inequality implies that $E_{D}$ is weakly
continuous, which is not proved yet. The proof of weak continuity follows
from the same arguments of the proof of property (vii) of $E_{dc}$. Equation
(\ref{eq:hashing}) also implies 
\begin{equation}
E_{dc}(\rho )=E_{D}(\rho ).  \label{eq:Conjecture1}
\end{equation}
The proof of Eq.~(\ref{eq:Conjecture1}) is simple and essentially the same
as that in \cite{HHHb}. The partial additivity and the weak monotonicity of $%
E_{D}$ \cite{HHHa,Horodecki} give 
\begin{eqnarray}
E_{D}(\rho ) =\lim_{n\rightarrow \infty }\frac{1}{n}E_{D}(\rho ^{\otimes
n})
&\geq &\lim_{n\rightarrow \infty }\sup_{\Lambda _{n}}\frac{1}{n}E_{D}\left(
\Lambda _{n}(\rho ^{\otimes n})\right)   \nonumber\\
&\geq &\lim_{n\rightarrow \infty }\sup_{\Lambda _{n}}\frac{1}{n}I_{B}\left(
\Lambda _{n}(\rho ^{\otimes n})\right) =E_{dc}(\rho ).
\label{eq:Conjecture2}
\end{eqnarray}
From Eqs.~(\ref{eq:EDEdcERinf}) and (\ref{eq:Conjecture2}), Eq.~(\ref
{eq:Conjecture1}) is obtained. This is a satisfactory result. It strengthens
the information-theoretic meaning of the distillable entanglement; namely, $%
E_{D}$ is the ultimate measure of resources for dense coding. Furthermore,
the optimal entanglement distillation seems to be the best strategy to
maximize the coherent information since it increases $S\left( {\rm Tr}%
_{A}\left( \Lambda _{n}(\rho ^{\otimes n})\right) \right) $ on one hand and
decreases $S\left( \Lambda _{n}(\rho ^{\otimes n})\right) $ on the other
hand while keeping the dimension of $\Lambda _{n}(\rho ^{\otimes n})$ as
large as possible. According to the above reasonings, it is most likely that 
$E_{dc}(\rho )=E_{D}(\rho )$. Unfortunately, the assumed inequality $%
E_{D}(\rho )\geq I_{B}(\rho )$, which is also a consequence of the equality $%
E_{dc}(\rho )=E_{D}(\rho )$ [Eq.~(\ref{eq:Conjecture1})], is not proven yet.
One of the possible counter-examples is {\it Example 1}. However, Rains has
conjectured that for any maximally correlated state $\rho $ both the PPT
distillable entanglement and the one-way distillable entanglement coincide 
\cite{Rains01}, so $E_{D}(\rho )=I_{B}(\rho )$. It should be noted that this
conjecture is also a consequence of the hypothetical hashing inequality.

In summary, in the light of the dense coding capacity optimized with respect
to LQCC, an asymptotic entanglement measure $E_{dc}$ for any bipartite
states was derived and its properties was investigated. Some examples of $%
E_{dc}$ with explicit forms were also given. Finally, it was argued that $%
E_{dc}$ coincides with the distillable entanglement. A possible
counter-example for this conjecture was also given.

The author would like to thank Andreas Winter for helpful comments.
This work was supported by CREST of Japan Science and Technology Corporation
(JST).


\begin{thebibliography}{99}

\bibitem{BW}  C. H. Bennett and S. J. Wiesner, Phys. Rev. Lett. {\bf 69},
2881 (1992).

\bibitem{BSST}  C. H. Bennett, P. W. Shor, J. A. Smolin, and A. V.
Thapliyal, Phys. Rev. Lett. {\bf 83}, 3081 (1999).

\bibitem{BPV}  S. Bose, M. B. Plenio, and V. Vedral, J. Mod. Opt. {\bf 47},
291 (2000).

\bibitem{Hiroshima}  T. Hiroshima, J. Phys. A {\bf 34}, 6907 (2001).

\bibitem{Bowen}  G. Bowen, Phys. Rev. A {\bf 63}, 022302 (2001).

\bibitem{HHHLT}  M. Horodecki, P. Horodecki, R. Horodecki, D. W. Leung, and
B. M. Terhal, Quantum Inf. Comp. {\bf 1}, 70 (2001).

\bibitem{Winter}  A. Winter, quant-ph/0108066.

\bibitem{Kholevo}  A. S. Kholevo, Probl. Peredachi Inf. {\bf 9}, 3 (1973)
[Probl. Inf. Transm. (USSR) {\bf 9}, 110 (1973)].

\bibitem{Holevo}  A. S. Holevo, IEEE Trans. Inf. Theory, {\bf 44}, 269
(1998).

\bibitem{SW}  B. Schumacher and M. Westmoreland, Phys. Rev. A {\bf 56}, 131
(1997).

\bibitem{PVP}  M. B. Plenio, S. Virmani, and P. Papadopoulos, J. Phys. A 
{\bf 33}, L193 (2000).

\bibitem{VPRK}  V. Vedral, M. B. Plenio, M. A. Rippin, and P. L. Knight,
Phys. Rev. Lett. {\bf 78}, 2275 (1997).

\bibitem{VP}  V. Vedral and M. B. Plenio, Phys. Rev. A {\bf 57}, 1619 (1998).

\bibitem{COMMENT} In a strict sense $\chi _{\max }^{*}(\rho )$ [Eq.~(\ref{eq:Edc})] is not a dense coding capacity because the protocol involves two-way classical communications.

\bibitem{BDSW}  C. H. Bennett, D. P. DiVincenzo, J. A. Smolin, and W. K.
Wooters, Phys. Rev. A {\bf 54}, 3824 (1996).

\bibitem{AEJPVD}  K. Audenaert, J. Eisert, E. Jan\'{e}, M. B. Plenio, S.
Virmani, and B. De Moor, Phys. Rev. Lett. {\bf 87}, 217902 (2001).

\bibitem{HHHa}  M. Horodecki, P. Horodecki, and R. Horodecki, Phys. Rev.
Lett. {\bf 84}, 2014 (2000).

\bibitem{HH}  M. Horodecki and P. Horodecki, Phys. Rev. A {\bf 59}, 4206
(1999).

\bibitem{Horodecki}  M. Horodecki, Quantum Inf. Comp. {\bf 1}, 3 (2001).

\bibitem{DHR}  M. J. Donald, M. Horodecki, and O. Rudolph, quant-ph/0105017.

\bibitem{Fannes}  M. Fannes, Commun. Math. Phys. {\bf 31}, 291 (1973).

\bibitem{Rains00}  E. M. Rains, Phys. Rev. A {\bf 60}, 179 (1999); {\bf 63},
019902(E) (2000).

\bibitem{Rains01}  E. M. Rains, IEEE Trans. Inf. Theory {\bf 47}, 2921
(2001).

\bibitem{Rains99}  E. M. Rains, Phys. Rev. A {\bf 60}, 173 (1999).

\bibitem{EFPPW}  J. Eisert, T. Felbinger, P. Papadopoulos, M. B. Plenio, and
M. Wilkens, Phys. Rev. Lett. {\bf 84}, 1611 (2000).

\bibitem{HHHb}  M. Horodecki, P. Horodecki, and R. Horodecki, Phys. Rev.
Lett. {\bf 85}, 433 (2000).

\end{thebibliography}
\end{document}